\documentstyle[aps,prl,epsf,prb,twocolumn]{revtex}
\begin{document}
\twocolumn[\hsize\textwidth\columnwidth\hsize\csname @twocolumnfalse\endcsname

 \draft
\title{The Decay Properties of the Finite Temperature Density Matrix in Metals}
\author{S. Goedecker}
 \address{ Max-Planck-Institut f\"{u}r Festk\"{o}rperforschung \\
 Heisenbergstr. 1, 70569 Stuttgart, Germany}
\maketitle
 
]

\begin{abstract}      
Using ordinary Fourier analysis, the asymptotic decay behavior of 
the density matrix $F({\bf r},{\bf r}')$ is derived for the 
case of a metal at a finite electronic temperature. 
An oscillatory behavior which is damped exponentially with increasing distance 
between ${\bf r}$ and ${\bf r}'$ is found. The decay rate is not only determined 
by the electronic temperature, but also by the Fermi energy. 
The theoretical predictions are confirmed by numerical simulations.
\end{abstract}

\bigskip
PACS: 71.90.+q \hspace{1cm} 71.55.Ak
\bigskip

\narrowtext
The decay behavior of the density matrix is a fundamental property of 
quantum mechanical systems since it determines the degree of locality~\cite{nearsight} of all 
relevant quantities. Experimentally it is well known, that most quantities 
such as bonding properties and energies or magnetic moments depend only on 
an environment comprising a few nearest neighbor shells. Locality is 
certainly much stronger in insulators, but is also applies in metals.
Recently there has also been renewed interest in these decay properties, 
since they determine whether and for what system size so called O(N) methods~\cite{ON}  
are faster than traditional electronic structure algorithms.
The decay behavior of the density matrix in insulators has been the subject 
of several studies. It was shown that the decay constant is related to 
the location of branch points of the band structure in the complex energy 
plane~\cite{french1,french2}. Since this is a quantity which is not 
accessible in ordinary electronic 
structure calculations, it would of course be highly desirable to relate 
the decay to features of the real part of the band structure. For the case 
of periodic solids arguments can therefore be found proposing a relation 
between the size of the gap and the decay behavior~\cite{kohn1,kohn2,kohn3}. Based on an analysis 
of the expansion of the density matrix in terms of Chebychev polynomials, this 
was recently confirmed~\cite{gordondecay} for arbitrary systems. 

In this paper the decay properties of the finite temperature density matrix 
in metals will be studied.
Contrary to a widespread belief that the decay constant is given 
by $\sqrt{ k_B T}$ where $k_B$ is Boltzmann's constant and $T$ the electronic temperature.
we find that the decay is as well determined by the Fermi level.

In the finite temperature formalism of Mermin~\cite{mermin} the finite temperature 
density matrix $F$ is given by
\begin{equation} \label{ftrans}
F({\bf r},{\bf r}') = \sum_n \int_{BZ} d{\bf k} \: 
    f \left( \frac{\epsilon_n({\bf k})-\mu}{k_B T} \right) 
    \Psi^*_{n,{\bf k}}({\bf r}) \Psi_{n,{\bf {\bf k}}}({\bf r}')
\end{equation}
where $\Psi_{n,{\bf k}}({\bf r}) = u_{n,{\bf k}}({\bf r}) e^{i{\bf k}({\bf r})} $ 
are the Bloch functions
associated with the wave vector $k$ and $n$ is the band index.
The function $ f(\frac{\epsilon_n({\bf k})-\mu}{k_b T})$ is the Fermi 
distribution for the band-structure energy $\epsilon_n({\bf k})$ and for the 
chemical potential $\mu$.
The integral extends over the Brillouin zone (BZ).

The cell periodic part of the 
Bloch functions $u_{n,{\bf k}}({\bf r})$ can be pulled out of the integral using the medium 
value theorem.  Denoting the medium value by $k_0({\bf r},{\bf r}')$
one obtains
\begin{eqnarray} 
F({\bf r},{\bf r}') & = & \sum_n 
  u^*_ {n,{\bf k}_0({\bf r},{\bf r}')}({\bf r}) u_{n,{\bf k}_0({\bf r},{\bf r}')}({\bf r}')  \\
& &\int_{BZ} d{\bf k} \: f\left( \frac{\epsilon_n({\bf k})-\mu}{k_B T} \right) 
     e^{i {\bf k}({\bf r}-{\bf r}')} \nonumber
\end{eqnarray}
The part we have taken out is periodic in both $k$ and $r$ and bounded from above 
since the wavefunctions $u_{n,{\bf k}}({\bf r})$ are normalizable. 
If $n$ refers to the lowest band 
the expression can also never vanish since $u_{n,{\bf k}}({\bf r})$  has no nodes. 
Hence the expression is 
bounded from below as well and it can therefore not influence the 
asymptotic behavior.
In the case of higher bands the expression can in principle vanish and 
might lead thus a faster asymptotic decay. If the $u_{n,{\bf k}}({\bf r})$ has for instance 
an $p_z$ like character, the cell periodic part can identically vanish for 
certain directions of ${\bf r}$ and for certain values of ${\bf r}'$. The expression 
can however certainly not vanish for all values of ${\bf r}$ and ${\bf r}'$.
Thus the conclusion that the cell periodic part does not 
change the overall asymptotic behavior remains valid.

In addition we will assume that the 
band structure can be approximated by the free electron band structure. 
The decay behavior of the metallic system is thus the same as the one of jellium
\begin{equation}
F({\bf r},{\bf r'}) =
      \int e^{i({\bf k}({\bf r}-{\bf r'}))}
      \: f \left( \frac{ \frac{1}{2} k^2 - \mu}{ T} \right) d{\bf k}
\end{equation}
Using elementary calculus this three-dimensional integral can be transformed into 
an one-dimensional one. Using integration by parts the Fermi distribution can be replaced 
by its derivative $f'$ and one obtains
\begin{equation}
F(s) =  - \frac{2 \pi}{T} \frac{1}{s} \frac{\partial}{\partial s} \frac{1}{s}
                       \frac{\partial}{\partial s} \int_{-\infty}^{\infty}
               e^{i k s} \: f' \left( \frac{ \frac{1}{2} k^2 - \mu}{T} \right) dk
\end{equation}
where $s = |{\bf r}-{\bf r'}|$.

To proceed further with an analytic evaluation of the resulting  
expression one has to approximate $f'$ by a replacement function. The basic idea is that 
as the temperature tends to zero the part coming from $f'$ becomes a delta function.
So in the following steps we will replace $f'$ by another representation of the 
delta function and we will use the rules which are valid for delta function 
for our sharply peaked finite temperature function as well even though it 
becomes a delta function only in the 
limit of zero temperature. A good choice for the replacement function is the following
\begin{equation}
f'(x) \rightarrow - \frac{sinh(a x)}{sinh(b x)}
\end{equation}
where $a = \frac{\pi}{4} \: tan \left( \frac{\pi}{8} \right)$ and
$b = \pi \: tan \left( \frac{\pi}{8} \right)$. The coefficients $a$ and $b$ were chosen 
such that the replacement function has the same local and global behavior as the 
original function.
The evaluation of the remaining integral then gives
\begin{equation} \label{jellap}
F(s) =  \frac{4 \pi \: c}{k_F s} \frac{\partial}{\partial s} \frac{1}{s}
          \frac{\partial}{\partial s} 
 \left[ \frac{cos(k_F s)} { 1+ \sqrt{2} cosh \left( - c \frac{k_B T}{k_F} s \right) } \right]
\end{equation}
where $c = 1+\sqrt{2}$ and where $k_F$ is the Fermi wave-vector related to the 
chemical potential by $\frac{1}{2} k_F^2 = \mu$.

\begin{figure}[ht]
     \begin{center}
      \setlength{\unitlength}{1cm}
       \begin{picture}( 6.,6.)           
        \put(-2.5,-1.){\includegraphics{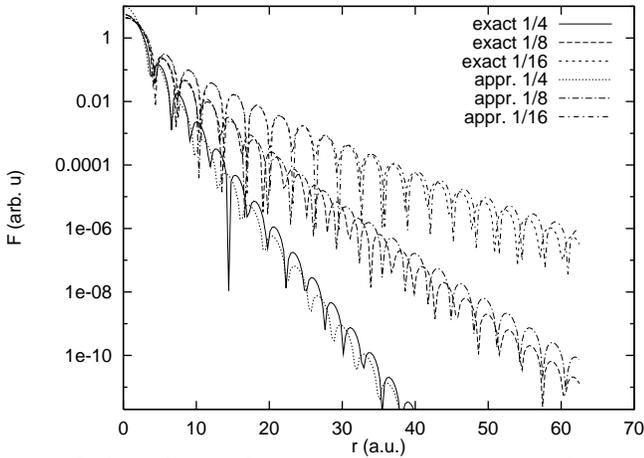}}   
       \end{picture}
       \caption{\label{jell} \it Comparison of the exact decay behavior of the
                density matrix and its approximation of Equation~\ref{jellap}.
                Three different thermal energies $k_B T$ of $1/4$, $1/8$ and $1/16$
                are plotted for $k_F=1$.}
      \end{center}
     \end{figure}

So we see two prominent features in the behavior of the density matrix.
It has oscillations whose wave-vector is given by the Fermi wave-vector $k_F$
and it is decaying exponentially, since for large values of $r$ the expression
$ \left( 1+ \sqrt{2} cosh \left( - c\frac{k_B T}{k_F} r \right) \right)^{-1}$
tends to $exp(- c \frac{k_B T}{k_F} r)$.
Despite the approximative character of Equation~\ref{jellap} one
obtains very good agreement in the jellium case 
between the exact density matrix obtained numerically and its
approximation (Equation~\ref{jellap}) over many orders of magnitude
as shown in Figure~\ref{jell}.

Let us note that all the approximations used in the derivation become exact as the 
temperature tends to zero and consequently we obtain the correct zero temperature 
result. 
\begin{equation} 
F(s) = - \frac{4 \pi k_F}{s^2}  \left(
          cos(s k_F) - \frac{sin(s k_F)}{s k_F} \right)
\end{equation}
This confirms the well known fact that the amplitude decays algebraically in the zero 
temperature case. More specifically a decay proportional to $r^{-2}$ is found.

Since the derivation was based on the assumption that the band structure is free 
electron like, the results may not be applicable in the case of metals 
which a a very complicated Fermi surface arising from a band-structure which 
is completely different from the free electron band structure.

I thank O. Gunnarson for his valuable comments, 
which considerably improved the paper.

\end{document}